\documentclass{article}
\usepackage[utf8]{inputenc}
\usepackage{relsize}
\usepackage{amsmath}
\usepackage{mathtools}
\usepackage{amssymb}
\usepackage{cite}
\usepackage{xspace}
\usepackage{comment}
\usepackage{authblk}

\newcommand{\bk}{{\textbf{k}}}
\newcommand{\EF}{\ensuremath{E_{\textrm{F}}}\TextOrMath{\@\xspace}{}}

\usepackage{geometry}
\title{Semiconductor nanodevices as a probe of strong electron correlations}
\author[1]{Pedro Vianez}
\author[2]{Oleksandr Tsyplyatyev}
\author[1]{Christopher Ford}
\affil[1]{Cavendish Laboratory, University of Cambridge, Cambridge, United Kingdom}
\affil[2]{Institut f\"ur Theoretische Physik, Universit\"at Frankfurt, Frankfurt, Germany}

\begin{document}

\maketitle

\begin{abstract}
Interactions between electrons in solids are often behind exciting novel effects such as ferromagnetism, antiferromagnetism and superconductivity. All these phenomena break away from the single-electron picture, instead having to take into account the collective, correlated behaviour of the system as a whole. In this chapter we look at how tunnelling spectroscopy can be used as the experimental tool of choice for probing correlation and interaction effects in one-dimensional (1D) electron systems. We start by introducing the Tomonaga-Luttinger Liquid (TLL) model, showing how it marks a clear departure from Fermi-liquid theory. We then present some early experimental results obtained using tunnelling devices and how they contributed to the decisive observation of both spin-charge separation and power-law behaviour. Other experimental techniques, such as photoemission and transport measurements, are also discussed. In the second half of the chapter we introduce two nonlinear models that are counterparts to the TLL theory, known as the mobile-impurity and the mode-hierarchy pictures, and present some of the most recent experimental evidence in support of both.
\end{abstract}


\newpage
\tableofcontents
\newpage
\section{The failure of Fermi-liquid theory}

Many-body interactions of electrons in three or two dimensions can, generally speaking, be treated using Fermi-liquid theory, proposed by Lev Landau in 1956 \cite{landau_the,altland_condensed_2010}. Landau showed that, despite the enormous strength of electron-electron interactions, these systems could be described as collective excitations that behave in almost every way very similarly to free electrons, but have different effective masses. In Fermi-liquid theory electrons can therefore be modelled as if being part of a non-interacting system.

Let us consider a so-called `normal' Fermi liquid. Here, the system is in a ground state when all interactions are switched off, the states being filled according to Fermi-Dirac statistics. This forms a well-defined Fermi surface at low temperatures. In the non-interacting regime, an excited state can then be represented by a filled Fermi circle or sphere, corresponding to an equilibrium state of zero momentum, together with an additional particle of momentum $p$. As interactions are switched on, conservation of the total momentum enforces conservation of the momentum $p$ of the same excited state, even as the additional particle moves across the system causing a perturbation. Landau proved that this separate entity---which he called a quasiparticle, composite of contributions from many interacting particles---could move around and only very weakly interact with its surroundings, essentially as if interactions were absent. 

Non-Fermi liquids mark the departure from this view. For an in-depth review of the field see \cite{schofield_non-fermi_1999}. In general however, there are two scenarios where Fermi-liquid theory usually fails:

\begin{enumerate}
    \item In higher dimensions, such as two- (2D) or three-dimensional (3D) systems, if the electron liquid is highly correlated, for instance when the energy of the Coulomb interaction greatly exceeds the kinetic energy;
    \item In lower dimensions, like one-dimensional (1D) wires, where spatial confinement alone dictates strong correlations regardless of the strength of the interactions in question.
\end{enumerate}

\begin{figure}[h]
  \includegraphics[width=0.6\textwidth]{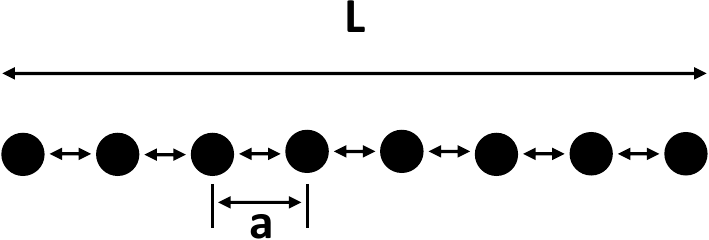}
  \centering
  \caption{Schematic representation of a one-dimensional electron system with density per unit length $N/L=1/a$, where $L$ is the length of the system and $a$ is the inter-particle spacing.}
  \label{fig:electrons_1D}
\end{figure}

In this chapter we focus on the latter and on how tunnelling spectroscopy can be used to experimentally probe such correlations. When confined to a 1D geometry (see Fig.\ \ref{fig:electrons_1D}), electrons cannot be modelled as Fermi quasiparticles, since each one is now strongly affected by its two neighbours. In other words, spatial confinement alone dictates that they are unable to go over, under or around each other, the only option left being to try and go through, were this not prevented by the diverging Coulomb repulsion at short distances. Hence, strong correlations arise.

One-dimensional wires mark a drastic departure from their higher-dimensional counterparts as, unlike the latter, there is no longer such a thing as locality of interactions. Instead, the motion of one electron causes a collective response of the whole system. This is why the single-particle properties, on which Fermi-liquid theory rests, break down completely. What emerges instead are the hydrodynamic modes of the Tomonaga-Luttinger liquid (TLL), with bosonic instead of fermionic statistics. We start this chapter by introducing the main ideas behind the TLL model, which replaces Fermi-liquid theory for electron liquids in one dimension, together with its main predictions.

\subsection{The Tomonaga-Luttinger model}

Many attempts have been made to describe strongly-correlated systems, especially in one and two dimensions. While at first sight, perturbation theory may seem like the obvious approach, it is important to note that any such theory would only work under the condition of weak interactions. Proposed by Tomonaga in 1950 \cite{tomonaga_remarks_1950} and later refined by Luttinger in 1963 \cite{luttinger_exactly_1963}, the Tomonaga-Luttinger-liquid model was introduced as an attempt to describe interacting one-dimensional electron systems at arbitrary interaction strength. The correct solution of the model was first obtained by Mattis and Lieb in 1965 \cite{mattis_exact_1965} and later improved by Luther and Peschel in 1974 \cite{luther_single-particle_1974}, before being finalised by Haldane in 1981 \cite{haldane_textquotesingleluttinger_1981}. The idea was that by using Bloch's theory of sound waves, electron excitations could be described using a Bose field, effectively laying the groundwork for what today is known as the method of bosonisation.

\begin{figure}[h]
  \includegraphics[width=\textwidth]{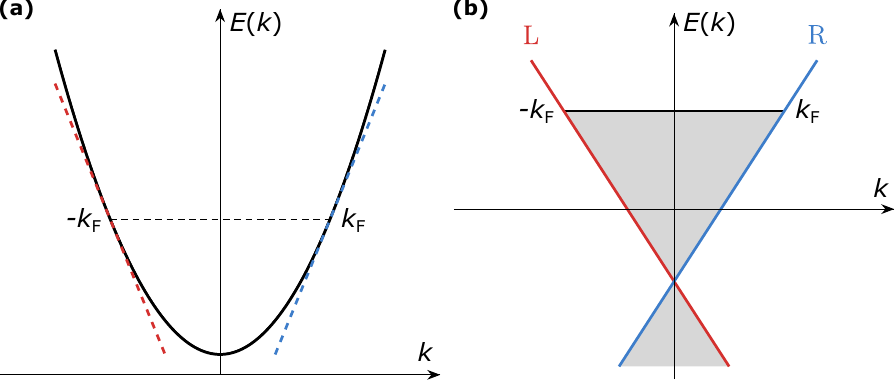}
  \centering
  \caption{The Tomonaga-Luttinger-liquid (TLL) model. (a) The original energy dispersion relation is shown in black. For  electrons in free space, this is  parabolic. According to the TLL picture, this can be approximated by a linear dispersion close to the Fermi points (red and blue dashed lines). (b) The linear dispersion used in the TLL model. Red and blue curves correspond, respectively, to left- and right-moving electrons, while the shaded area marks the filled Dirac sea below the Fermi energy. (From \cite{Jin_2020}, \textit{reprinted with permission}.)}
  \label{fig:luttinger_approximation}
\end{figure}

\sloppy The Tomonaga-Luttinger model is used to describe highly-correlated interacting one-dimensional systems. It was initially developed as a simple spinless theory but has since been extended to include spin \cite{giamarchi_quantum_2003}. Although exactly solvable, it starts from two key assumptions, that the dispersion is linear near the Fermi points and that the 1D system is infinitely long, see Fig.\ \ref{fig:luttinger_approximation}. We know, however, that the dispersion of free electrons, given by $E_{k}=\hbar^2k^2/\left(2m^*\right)$, is parabolic.
Nevertheless, when low-energy excitations occur, electrons can only move from just below to just above the Fermi energy and so, under these conditions, even though the true dispersion is in fact slightly curved (i.e.\ nonlinear), one can argue approximate linearity. Similarly, the assumption of infinitely long wires can be justified as long as end effects are not expected to play a significant role. A final assumption is that the wire is infinitely narrow, i.e.\ that the subband spacing is infinite, so that there is no mixing with higher 1D subbands. This makes it hard to be sure whether the theory is applicable to any particular experimental observation (where subband spacing is never huge compared with the interaction energy).

In three or two dimensions, low-energy excitations with arbitrarily low momenta can occur \cite{giamarchi_quantum_2003}. This is because in both 3D and 2D the Fermi surface in $k$-space can be modelled as a sphere or a circle, respectively, with radius $k_{\textrm{F}}$, and electrons are able to occupy any state within it. In 1D systems, however, this is no longer the case, as the Fermi surface is now not only discontinuous but also consists of only two specific points, at $k_{\textrm{F}}$ and $-k_{\textrm{F}}$. Within the TLL assumptions, one can write the Hamiltonian of a one-dimensional interacting electron system as
\begin{equation}
    \hat H=\hat H_{0}+\hat H_{\textrm{int}}.
    \label{eq:Hamiltonian_total}
\end{equation}
Here, $\hat H_{0}$ represents the kinetic energy, while $\hat H_{\textrm{int}}$ refers to the interaction energy of the system. They can be written respectively as
\begin{equation}
    \hat H_{0}=\sum_{k,\alpha=\pm1}\upsilon_{\textrm{F}}(\alpha k-k_{\textrm{F}})\hat a_{k,\alpha}^{\dagger}\hat a_{k,\alpha},
    \label{Hamiltonian_kinetic}
\end{equation}
where $\upsilon_{\textrm{F}}=\hbar k_{\textrm{F}}/m$ is the Fermi velocity, $\hat a_{k,\alpha}^{\dagger}$ and $\hat a_{k,\alpha}$ are the creation and annihilation operators at the right ($\alpha=1$) or left ($\alpha=-1$) branches of the dispersion, and
\begin{equation}
    \hat H_{\textrm{int.}}=\frac{1}{2L}\sum_{q\neq0}\upsilon_{q}\hat\rho_{-q}\hat\rho_{q},
    \label{eq:Hamiltonian_interaction}
\end{equation}
where $\hat\rho_{q}$ is the electron density fluctuation operator and $\upsilon_{q}$ is the Fourier transform of the interaction potential. Together, $\hat H$ accounts for particles moving in either of the right- ($R$) and left- ($L$) branches, as well as for inter- and intra-branch interactions \cite{giuliani_quantum_2005}. 

\begin{figure}[h]
  \includegraphics[width=\textwidth]{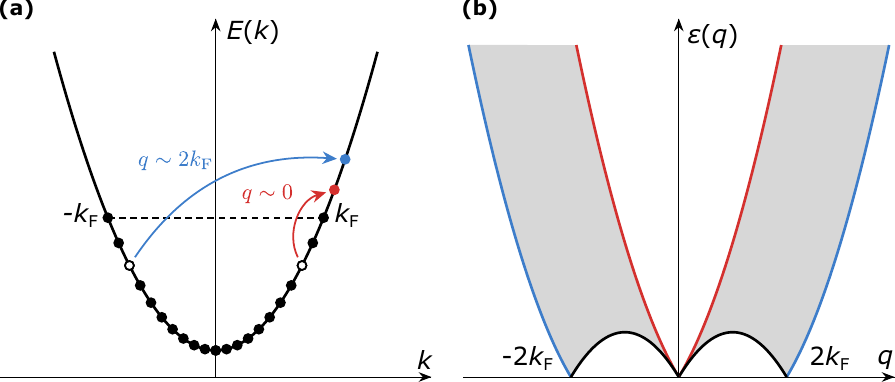}
  \centering
  \caption{ Excitation Spectrum for 1D fermions. (a) Single-particle excitations constituting forward (red) and backward (blue) scattering for small and large momentum change across the Fermi points, respectively. Occupied states are represented by filled dots and holes by empty circles. (b) Electron-hole pair spectrum. Only shaded areas are energetically accessible. The forbidden region indicated by the gap is unique to the 1D geometry and is not present at higher dimensions. The position of the spectral edge, separating the accessible and forbidden regions, is marked by the black curve. It can be observed directly, for instance, by neutron scattering in some magnetic materials  \cite{Lake05,Mourigal13,Lake13}. (From \cite{Jin_2020}, \textit{reprinted with permission}.)}
  \label{fig:right_left_movers}
\end{figure}

Generally speaking, it is $\hat H_{\textrm{int.}}$ that characterises the type of interactions. For a 1D electron system with spacing $a$, these will be Coulomb in nature and of two types (see Fig.\ \ref{fig:right_left_movers}a): if an electron is excited within the same branch (left or right) from a state just below the Fermi energy, then only a small change in momentum is required, $q\sim 0$, with the direction of motion remaining unchanged. This is called forward scattering. On the other hand, if there is an exchange involving both branches, then a larger change in momentum will be necessary, $q\sim 2k_{\textrm{F}}$, with the direction of motion now being reversed. This constitutes backward scattering. All of these processes appear naturally from 
the density fluctuation operator when the particles in Fig.\ \ref{fig:right_left_movers}a are split into the left and the right moving subbands around the $\pm k_\textrm{F}$ points,
\begin{equation}
     \hat\rho_{q}\simeq
    \begin{cases}
      \sum\limits_{k}(\hat a_{k-q,\textrm{R}}^{\dagger}\hat a_{k,\textrm{R}}^{}+\hat a_{k-q,\textrm{L}}^{\dagger}\hat a_{k,\textrm{L}}^{}) & \text{for $|q|\sim 0$},\\
      \sum\limits_{k}\hat a_{k-q,\textrm{L}}^{\dagger}\hat a_{k,\textrm{R}}^{} & \text{for $q\sim 2k_{\textrm{F}}$}, \\
      \sum\limits_{k}\hat a_{k-q,\textrm{R}}^{\dagger}\hat a_{k,\textrm{L}}^{} & \text{for $q\sim -2k_{\textrm{F}}$.}
    \end{cases}  
    \label{eq:density_operator}
\end{equation}

By substituting Eq. (\ref{eq:density_operator}) into Eq. (\ref{eq:Hamiltonian_interaction}), the interaction term in the Hamiltonian can be divided into two terms depending on the direction of motion of the interacting electrons. We then take the long-wavelength limit, \emph{i.e.} by considering a system of infinite length or, in other words, by assuming arbitrarily high excitation energies, finding that the total TLL Hamiltonian is given by
\begin{equation}
\begin{split}
    \hat H_{\textrm{TLL}}&= \hat  H_{0}+\frac{1}{2L}\sum\limits_{q\neq0}V_{1}(q)(\hat\rho_{-q,\textrm{R}}\,\hat\rho_{q,\textrm{R}}+\hat\rho_{-q,\textrm{L}}\,\hat\rho_{q,\textrm{L}}) \\& +\frac{1}{2L}\sum\limits_{q\neq0}V_{2}(q)(\hat\rho_{-q,\textrm{R}}\,\hat\rho_{q,\textrm{L}}+\hat\rho_{-q,\textrm{L}}\,\hat\rho_{q,\textrm{R}}),
\end{split}
\label{eq:Hamiltonian_spinless}
\end{equation}
where $V_{1}(q)=\upsilon_{q}$ and $V_{2}(q)=\upsilon_{q}-\upsilon_{2k_{\textrm{F}}}$. Note that the choice of $V_1$ and $V_2$ is, in general, completely arbitrary and highly dependent on the interaction process in question. Attempts to verify their form have been made by Dash \textit{et al.} \cite{dash_does_2001} assuming a Gaussian-dependence, by Creffield \textit{et al.} \cite{creffield_spectral_1995,creffield_spin_2001} using Monte-Carlo simulations, and by Ha\"usler \textit{et al.} \cite{hausler_tomonaga-luttinger_2002} and Matveev \textit{et al.} \cite{matveev_conductance_2004} analytically.

It can also be shown that the TLL Hamiltonian is diagonalisable if written in a bosonic basis. This means that every operator, including the fermionic creation and annihilation operators, can be represented in terms of boson operators together with operators that raise or lower the \textit{total} number of particles. This is in stark contrast with the representation used so far, which assumed action on individual electrons and not the system as a whole. This is, however, not surprising. As argued before, in 1D systems, collective response replaces individual behaviour.

A full review of the bosonisation method can be found in Apostol \cite{apostol_bosonisation_1983} or in von Delft and Schoelle \cite{vonDelft98} as well as in several textbooks \cite{altland_condensed_2010,giamarchi_quantum_2003}. The main result is that a one-dimensional system, following the TLL assumptions, is equivalent to a system of independent massless bosons, where the dispersion  is given by $\omega_{q}=v|q|$, see Fig.\ \ref{fig:right_left_movers}b, with the velocity
\begin{equation}
    v=\lim_{q\rightarrow 0}\sqrt{\left|\upsilon_{\textrm{F}}+\frac{V_1(q)}{2\pi\hbar}\right|^2-\left|\frac{V_2(q)}{2\pi\hbar}\right|^2}.
    \label{eq:spinless_dispersion}
\end{equation}
Rewriting $v=v_{\textrm{F}}/K$ in eq.\ \ref{eq:spinless_dispersion} allows one to extract the TLL parameter $K$ as
\begin{equation}
    K=\lim_{q\rightarrow 0}\frac{1}{\sqrt{\left|1+\frac{V_1(q)}{2\pi\hbar\upsilon_{\textrm{F}}}\right|^2-\left|\frac{V_2(q)}{2\pi\hbar\upsilon_{\textrm{F}}}\right|^2}}.
    \label{eq:spinless_coupling}
\end{equation}
Here, $K$ encapsulates both the interaction potentials $V_1(q)$ and $V_2(q)$ as well as the Fermi velocity $\upsilon_{\textrm{F}}$. Note that all of these parameters are independent of temperature.

\subsection{Spinful Tomonaga-Luttinger liquids}

Any model attempting to describe electron behaviour must account for spin. This plays an important role when considering the interaction process given by forward and backward scattering, as was shown by Luther \textit{et al.} \cite{luther_backward_1974,luther_single-particle_1974}, since both parallel and anti-parallel configurations are now possible. One can generalise both the kinetic $\hat H_0$ and the interaction $\hat H_{\textrm{int.}}$ terms to include spin, see section 9.12 of \cite{giuliani_quantum_2005}. The Hamiltonian for the spinful Tomonaga-Luttinger liquid (sTLL) then becomes
\begin{equation}
\begin{split}
    \hat H_{\textrm{sTLL}}&= \hat  H_{0}+\frac{1}{2L}\sum\limits_{q\neq0}V_{1}(q)\left(\hat\rho_{-q,\textrm{R}}\,\hat\rho_{q,\textrm{R}}+\hat\rho_{-q,\textrm{L}}\,\hat\rho_{q,\textrm{L}}\right) \\& +\frac{1}{2L}\sum\limits_{q\neq0,\sigma}\left[V_{2}(q)-V_1(q)\right]\left(\hat\rho_{-q\sigma,\textrm{R}}\,\hat\rho_{q\sigma,\textrm{L}}+\hat\rho_{-q\sigma,\textrm{L}}\,\hat\rho_{q\sigma,\textrm{R}}\right),
\end{split}
\label{eq:Hamiltonian_spinful}
\end{equation}
where $\hat \rho_{q,\alpha}\equiv\hat \rho_{q\uparrow,\alpha}+\hat \rho_{q\downarrow,\alpha}$ is the electron density operator and $\sigma=\uparrow, \downarrow$ is the spin index. 

Note the similarities between the spinless (Eq.\ \ref{eq:Hamiltonian_spinless}) and the spinful (Eq.\ \ref{eq:Hamiltonian_spinful}) Hamiltonians, the only difference being the potential prefactor of the last term. Therefore, the bosonisation method remains applicable as long as both bosonic operators and the (spin-)density fluctuation operator are also modified to include spin. 

One of the most striking features of the spinful TLL is what came to be known as spin-charge separation. Once $\hat H_{\textrm{sTLL}}$ is rewritten in terms of the spinful bosonic operators, one sees that it decouples into two independent Hamiltonians,
\begin{equation}
    \hat H_{\textrm{sTLL}}=\hat H_{\textrm{C}} + \hat H_{\textrm{S}},
\end{equation}
which can be associated with charge and spin modes, respectively. This result is analogous to the one discussed in the spinless case. Now, however, one has two independent collective excitations given by the characteristic frequencies $\omega_q^{\textrm{C}}=|q|v^{\textrm{C}}$ and $\omega_q^{\textrm{S}}=|q|v^{\textrm{S}}$, where
\begin{equation}
    v^{\textrm{C}}=\lim_{q\rightarrow 0}\sqrt{\left(\upsilon_{\textrm{F}}+\frac{V_1(q)}{\pi\hbar}\right)^2-\left(\frac{V_1(q)+V_2(q)}{2\pi\hbar}\right)^2}
\end{equation}
is the velocity of the charge-density wave (CDW) and
 \begin{equation}
     v^{\textrm{S}}=\lim_{q\rightarrow 0}\sqrt{\upsilon_{\textrm{F}}^2-\left(\frac{V_1(q)-V_2(q)}{2\pi\hbar}\right)^2}
 \end{equation}
is the velocity of the spin-density wave (SDW). These are, generally speaking, different.

\begin{figure}[h]
  \includegraphics[width=\textwidth]{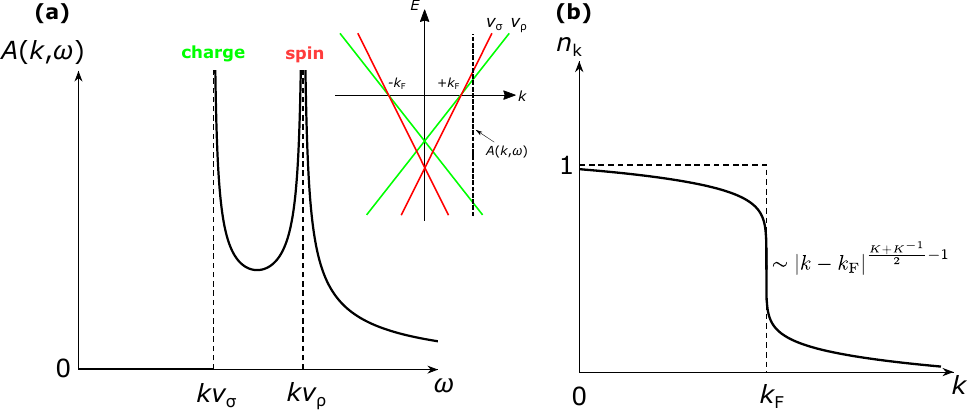}
  \centering
  \caption{ Spectral function of the spinful TLL and occupation numbers. (a) Spectral function $A(k,\omega)$ of a spinful TLL, evaluated at fixed $k$, showing charge and spin power-law singularities at different energies \cite{Voit93,Schoenhammer92}. (From \cite{schofield_non-fermi_1999}, \textit{reprinted with permission}.) Inset: Elementary excitations of a spinful TLL with $v_{\rho}\neq v_{\sigma}$. Green and red lines mark charge- and spin-type modes, respectively. (b) Occupation numbers $n_k$ within the TLL theory without spin. (From \cite{Jin_2020}, \textit{reprinted with permission}.)
  }
  \label{fig:spin_charge_power_law}
\end{figure}

Similarly to the spinless scenario, one can also define two interaction parameters for the spinful TLL, $K_{\rho,\sigma}$ given by $v_\rho\equiv v^{\textrm{C}}=v_{\textrm{F}}/K_\rho$ and $v_\sigma\equiv v^{\textrm{S}}=v_{\textrm{F}}/K_\sigma$ for the charge and spin modes, respectively. A spinful TLL is therefore fully determined by four parameters, $\upsilon_\rho$, $\upsilon_\sigma$, $K_\rho$ and $K_\sigma$, compared to just two, $\upsilon$ and $K$, in the spinless scenario.

\subsection{Spectral functions and power-law behaviour}

A rigorous way of describing single-particle properties in a many-body system can be given via the spectral function $A(k,\omega)$. Formally, it represents the probability density of increasing or decreasing the energy of an $N$-particle system by adding or removing a single particle from a state characterised by a momentum $k$ and energy $\omega$, see an example in Fig.\ \ref{fig:spin_charge_power_law}a for a spinful TLL. For comparison, in the absence of interactions, $A(k,\omega)$ can be shown \cite{giuliani_quantum_2005} to take the form of a $\delta$-function in frequency, $A(k,\omega)\propto\delta(\omega-E_k)$, \emph{i.e.} adding or removing a particle to/from a plane-wave state always results in an exact eigenstate of the system. Here, $E_k=\hbar^2\left(k_x^2+k_y^2\right)/\left(2m\right)$ for a 2D system.

In order to account for broadening arising from disorder, the spectral function can be convolved with a Lorentzian function. Following from the notation used by Kardynal \textit{et al.} \cite{kardynal_magnetotunneling_1997}, we have
\begin{equation}
    A_{\textrm{1(2)}}\left(k_x,k_y,k_x^{'},n;\mu\right)=\frac{\Gamma/\pi}{\left[\Gamma^2+\left(\mu-\xi_{1(2)}\right)^2\right]},
\label{eq:spectral_functions}
\end{equation}
where $\Gamma$ is the spectral line width associated with the single-particle disorder broadening, $\mu$ is the chemical potential,  and $\xi_{1(2)}$ are the energy spectra of the system, given by
\begin{align}
\xi_1=E_n+ \frac{\hbar^2{k^\prime}_x^{2}}{2m^{*}}+E_{0,1},\\ \xi_2=\frac{\hbar^2(k_x^2+k_y^2)}{2m^{*}}+E_{0,2}
\end{align}
with $E_n=(n-1/2)\hbar\omega_0$ being the energy of the \textit{n}th subband and $E_{0,1(2)}$ denoting the bottom of the 2D subband in the top and bottom wells, respectively. 

Experimentally however, most 1D systems are not \textit{perfect} 1D systems, \emph{i.e.} infinitely narrow, having instead a finite transverse extension. This can be accounted for by treating the confinement potential as being parabolic, with electron wave functions in the transverse direction given by the eigenstates of the quantum harmonic oscillator, as is the case in semiconducting quantum wires. Consequently, $A_1\left(\textbf{k},n;\mu\right)$ must be modified to account for a harmonic dependence along the $y$-direction, resulting in energy levels known as 1D subbands and given by
\begin{equation}
    A_{1,\textrm{non-int.}}\left(k^{\prime}_x,k_y,E,\Gamma\right)=\sum_n\frac{\Gamma}{\pi}\frac{H_n\left(k_ya\right)e^{-\left(k_ya\right)^2}}{\Gamma^2+\left(\mu-\xi_1\right)^2},
\label{eq:spectral_1D_nonint}
\end{equation}
where $a=m_{\textrm{1D}}\omega/\hbar$ is the finite width of the wire and $H_n(x)$ are the Hermite polynomials.

One can also find the momentum distribution function $n_k$ of the system in a similar fashion, as it is related to the spectral function via the real-space Green's function, see Fig.\ \ref{fig:spin_charge_power_law}b. A full analysis with the Green's function formalism is outside the scope of the review but can be found in \cite{abrikosov_methods_1976,giuliani_quantum_2005}. Most importantly, however, at $T=0$\,K and using Eq. (\ref{eq:spectral_1D_nonint}), $n(k)$ takes the following form:
\begin{equation}
    n(k)\propto|k-k_{\textrm{F}}|^{\frac{K+K^{-1}}{2}-1}.
\end{equation}
Note that, at $k=k_{\textrm{F}}$, $n(k)$ exhibits power-law behaviour with exponent $\alpha=(K+K^{-1})/2-1$. There is then a natural connection between the interaction parameter $K$, derived previously in the context of a spinless TLL, and correlation functions like $n(k)$ \cite{voit_charge-spin_1993}. As we will see in the next sections, early experimental studies of TLL-type behaviour focused exactly on this property, by trying to establish the existence of an interaction parameter $K$ by looking for power-law behaviour in the density of states around the Fermi points.

\section{Early results on Tomonaga-Luttinger-liquid behaviour}

The basic principle behind any spectroscopy technique is the measurement of the spectral function $A(k,\omega)$ which, for a 1D system, combines all the relevant information about the energy and momentum of the elementary excitations, both in the hole---when a particle is removed---and particle---when it is added---sectors, for arbitrary interaction strengths. From the solutions of the spinful TLL model we know that the spectral function of an interacting 1D electronic system is double-peaked, leading to spin-charge separation, and that the height of those peaks follows a power law in both temperature $T$ and source-drain bias $V_{\textrm{dc}}$. 

Early experimental work on verifying the TLL model focused on establishing this power-law nature as well as observing the separation between the spin and charge modes. It can be grouped into three categories following the experimental technique used and includes photoemission spectroscopy, momentum-resolved tunnelling spectroscopy and transport measurements. The range of systems utilised is large and their claims to be a 1D system vary significantly. Nevertheless, important progress has been made using GaAs quantum wires \cite{auslaender_spin-charge_2005,jompol_probing_2009}, carbon nanotubes \cite{bockrath_luttinger-liquid_1999,ishii_direct_2003,kim_tomonaga-luttinger_2007}, the high $T_{\textrm{c}}$ superconductor SrCuO$_{\textrm{2}}$ \cite{kim_observation_1996}, the 1D metal Li$_{\textrm{0.9}}$Mo$_{\textrm{6}}$O$_{\textrm{17}}$ \cite{gweon_generalized_2003}, 1D organic conductors \cite{zwick_band_1998,pouget_x_1976} and quantum-Hall edge states \cite{chang_chiral_2003,grayson_continuum_1998} to name but a few.

\subsection{Photoemission spectroscopy}

The most direct way of measuring a system's spectral function is by photoemission spectroscopy. This is a difficult experiment to do with high resolution and as such had limited applicability when it came to measuring if 1D systems truly behaved as given by the TLL model.

\begin{figure}[h]
  \includegraphics[width=\textwidth]{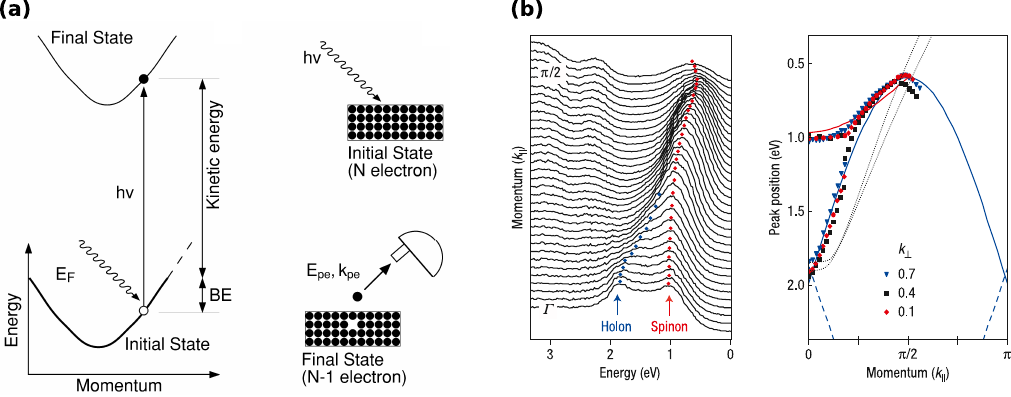}
  \centering
  \caption{(a) Schematic representation of the direct photoemission spectroscopy, used in mapping the band structure of a system. (From \cite{kim_spincharge_2001}, \textit{reprinted with permission}.) (b) Left: ARPES data showing the many-body spectrum of  $\textrm{SrCuO}_2$, where a two-peak spinon-holon structure can be seen, as predicted by the spinful Tomonaga-Luttinger model. Right: Comparison between experimental (symbols) and theoretical (solid and dashed lines) dispersions. (From \cite{kim_distinct_2006}, \textit{reprinted with permission}.)}
  \label{fig:Kim_merged}
\end{figure}

The principle of operation is, however, relatively simple, see Fig.\ \ref{fig:Kim_merged}a. An incident photon $E_{\gamma}=\hbar\omega$ of known energy causes the transition of an electron from an initial band state at energy $E_{\textrm{i}}$ to a final, plane-wave, state at energy $E_{\textrm{f}}$ above the vacuum energy, where $E_{\textrm{f}}=\hbar^2 k_{\textrm{f}}^2/\left(2m\right)=E_{\textrm{i}}+\hbar\omega-\phi$ and $\phi$ is the work function of the material. This photo-emitted electron leaves the crystal and can be collected by a detector. This gives the emission intensity $I(k,\omega)$, which can be related to the spectral function $A(k,\omega)$ by
\begin{equation}
    I(k,\omega)=M(k)A(k,\omega).
\end{equation}
Here, $M(k)$ is the photoemission matrix element involving the initial and final states of the emitted photoelectrons. By mapping the system's response of photo-emitted electron energies as a function of analyser angle, information about the band structure can be obtained. For a review of angle-resolved photoemission spectroscopy (ARPES) experiments, see  \cite{shen_electronic_1995}.

However, this technique is limited on two fronts. First, it assumes the 1D system is relatively close to the surface on which the light is incident. Anisotropic layered materials are a particularly good example of this, where there is little dispersion of the electronic bands perpendicular to the surface, making the data also easier to interpret. However, only certain classes of materials, such as specific high-temperature superconductors, fall into this category. Second, by its very nature, only occupied states can be probed. It is in principle possible to map unoccupied bands by inserting an electron of known energy into an empty band state and measuring the outgoing, ejected photon (this is known as \textit{inverse} photoemission spectroscopy), but little work has been carried out in the 1D field using this technique.

Photoemission spectroscopy was first applied by both Kim \textit{et al.} \cite{kim_observation_1996,kim_spincharge_2001,kim_distinct_2006} and Fujisawa \textit{et al.} \cite{fujisawa_angle-resolved_1998,fujisawa_spin-charge_1998} in studying the high-temperature superconductor $\textrm{SrCuO}_{\textrm{2}}$. Here, a good agreement to the experimental data was obtained by solving for the band structure using the Hubbard model, with the broadness of the observed peak being interpreted as resulting from the separation of the spinon and holon modes, see Fig.\ \ref{fig:Kim_merged}b. This work provided early evidence of spin-charge separation in a 1D Mott insulator which, although not metallic, still behaved as a 1D system as described by the TLL model.  

Another experiment using photoemission spectroscopy was carried out by Gweon \textit{et al.} \cite{gweon_generalized_2003} using the quasi-1D metal $\textrm{Li}_{\textrm{0.9}}\textrm{Mo}_{\textrm{6}}\textrm{O}_{\textrm{17}}$. Here, results similar to those shown above were also obtained, \emph{i.e.} spectral separation between the spinon and holon modes in the observed dispersion. From the spectral functions the velocities of each mode were also estimated, giving $v_{\rho}/v_{\sigma}=2$. Their experiment also yielded good theoretical agreement, with a power-law exponent of about $\alpha=0.9$, characteristic of a system with strong backscattering. 

\subsection{Transport measurements}\label{sec.transport}

Experimental studies on the TLL power-law behaviour were also performed by Bockrath \textit{et al.}  \cite{bockrath_luttinger-liquid_1999} and Yao \textit{et al.} \cite{yao_carbon_1999} using carbon nanotubes. Their experiments consisted of measuring electrical transport along single-walled carbon nanotubes (SWNTs) with different contact geometries, either bulk or end. This method relied on approximating the Luttinger-liquid parameter as
\begin{equation}
    K_{\rho}=\left[1+\frac{2U}{\Delta}\right]^{-1/2},
\end{equation}
where $U$ is the charging energy of the nanotube and $\Delta$ is the single-particle level spacing. Bockrath \textit{et al.} showed that, in a 1D system, the conductance along the system is suppressed at zero bias, vanishing as a power-law dependence with temperature $T$ for small biases $|eV|\ll k_{\textrm{B}}T$,
\begin{equation}
G(T)\propto T^\alpha,
\end{equation}
while with source-drain bias $V_{\textrm{dc}}$ at larger biases $|eV|\gg k_{\textrm{B}}T$,
\begin{equation}
    \textrm{d}I/\textrm{d}V\propto V^\alpha,
\end{equation}
as predicted by the TLL model. Here, $\alpha$ depends on the number of 1D channels and on whether the tunnelling is happening into the bulk of the system or whether it is affected by the ends. This can be found from the TLL theory by calculating the momentum distribution function once the spectral function is known. In SWNTs, depending on the contact geometry, the power-law exponent $\alpha$ can then be related to $K_{\rho}$ as either
\begin{equation}
    \alpha_{\textrm{end}}=(K_{\rho}^{-1}-1)/4
\end{equation}
or
\begin{equation}
    \alpha_{\textrm{bulk}}=(K_{\rho}^{-1}+K_{\rho}-2)/8
\end{equation}
as different contact geometries slightly change the prominence of the long-range Coulomb interactions.

In the experiment, sometimes individual nanotubes were placed on top of two metallic contacts, perturbing the nanotube only weakly and therefore yielding a long 1D system and hence `bulk' tunnelling. Alternatively, contact metal was deposited on top of the nanotube, yielding a short 1D system in which any tunnelling excitation would propagate to the ends and be affected by the boundary condition there, producing `end' tunnelling. This is determined by comparing the energy of the tunneling electron to $\Delta E = 2\hbar v_\textrm{F1D}/(K_{\rho}L)$, which is related to the inverse time scale for the TLL quasi-particles to travel to the end of a wire of length $L$ \cite{tserkovnyak_interference_2003,kane_transmission_1992}. For $k_\textrm{B}T$, $eV_\textrm{dc} \gg \Delta E$, the ends are unimportant.
 
\begin{figure}[h]
  \includegraphics[width=0.8\textwidth]{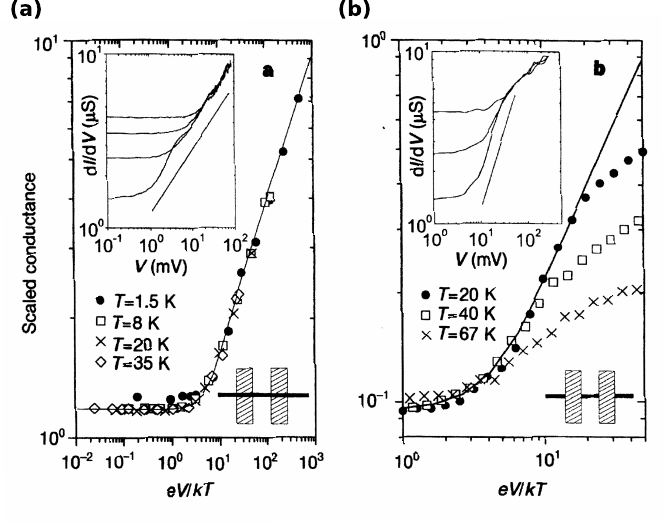}
  \centering
  \caption{Differential conductance $\textrm{d}I/\textrm{d}V$ taken at various temperatures for both (a) bulk- and (b) end-contacted carbon nanotubes. Insets show $\textrm{d}I/\textrm{d}V$ on a log-log plot where power-law behaviour, $\textrm{d}I/\textrm{d}V\propto T^\alpha$, can be seen in both geometries (see straight line). Main panels show the same data collapsed onto a single curve after using the scaling relation described in Eq. (\ref{eq:scaling_conductance}). The extracted values of $\gamma$ were (a) $\gamma=0.46$ and (b) $\gamma=0.63$. (From \cite{bockrath_luttinger-liquid_1999}, \textit{reprinted with permission}.)}
  \label{fig:Bockrath_Luttinger_1999}
\end{figure}

Conductance measurements as a function of temperature $T$ and source-drain bias $V_{\textrm{dc}}$ were performed at zero magnetic field, resulting in $\alpha_{\textrm{end}}\approx 0.6$ and $\alpha_{\textrm{bulk}}\approx 0.3$, in good agreement with the theoretical predictions for SWNTs, $\alpha_{\textrm{end}}(\textrm{theory})=0.65$ and $\alpha_{\textrm{bulk}}(\textrm{theory})=0.24$  \cite{kane_coulomb_1997,egger_effective_1997}, see insets in Fig.\ \ref{fig:Bockrath_Luttinger_1999}.

While different contact geometries translated into different transport mechanisms, both systems were observed to still follow a power law in temperature once the data was rescaled using the universal relation 
\begin{equation}
    \frac{\textrm{d}I}{\textrm{d}V}=AT^{\alpha}\textrm{cosh}\left(\gamma\frac{\textit{eV}}{2k_{\textrm{B}}T}\right)\left|\Gamma\left(\frac{1+\alpha}{2}+\gamma\frac{\textit{{\rm i}eV}}{2\pi k_{\textrm{B}}T}\right)\right|^2,
    \label{eq:scaling_conductance}
\end{equation}
where $\Gamma(x)$ is the gamma function, $\gamma$ takes into account the voltage drops at the two tunnel junctions, and $A$ is an arbitrary constant (see Fig.\ \ref{fig:Bockrath_Luttinger_1999}). Note that this result assumed that the leads were at $T=0$\,K. From here, the bulk-contacted nanotubes were observed to follow $\gamma_{\textrm{bulk}}\approx0.46$, while end-contacted nanotubes gave $\gamma_{\textrm{end}}\approx0.63$, in good agreement with the theoretical predictions of Fisher and Kane \cite{kane_transport_1992}, where $\gamma_{\textrm{bulk}}(\textrm{theory})=0.50\pm0.1$ and $\gamma_{\textrm{end}}(\textrm{theory})=0.60\pm0.1$. This was the first experimental evidence that electrons in a metallic nanotube behaved as a TLL.

\subsection{Magnetotunnelling spectroscopy}

Following Haldane's suggestion in 1981 and after initial measurements done on the 1D spectral function by Kardynal \textit{et al.} \cite{kardynal_direct_1996}, Barnes proposed in 1999 a novel method for probing TLL behaviour \cite{altland_magnetotunneling_1999}.

\begin{figure}[h]
  \includegraphics[width=\textwidth]{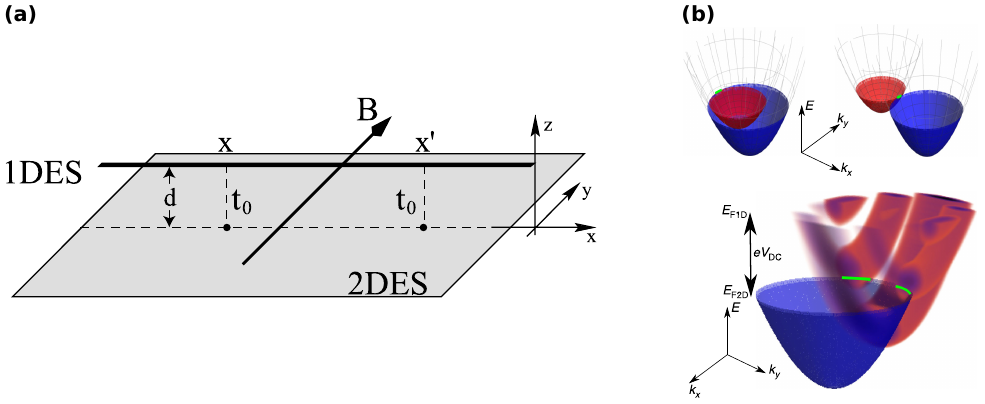}
  \centering
  \caption{(a) Device configuration for mapping the dispersion of a 1D system by measuring tunnelling of electrons from it to a nearby 2DES. (From \cite{altland_magnetotunneling_1999}, \textit{reprinted with permission}.) (b) Magnetotunnelling spectroscopy. Top panels show the overlap of the spectral functions of two 2D systems. Magnetic field $B$ displaces the blue paraboloid to the right, making the Fermi circles touch from the inside (top left) or from the outside (top right). This allows the states of the red paraboloid near $-k_{\textrm{F}}$ and $k_{\textrm{F}}$ to be probed at the Fermi energy $E_{\textrm{F}}$. Bottom panel shows tunnelling between a 2D (blue) and a 1D (red) system at finite bias $V_{\textrm{dc}}$. Multiple 1D subbands can be probed by varying both $B$ and $V_{\textrm{dc}}$. (From \cite{jin_momentum-dependent_2019}, \textit{reprinted with permission}.)}
  \label{fig:tunnelling_spectroscopy}
\end{figure}

The idea was to measure the tunnelling conductance between a quantum wire and a parallel two-dimensional electron system (2DES) as a function of both the potential difference between them, $V$, and an in-plane magnetic field, $B$, see Fig.\ \ref{fig:tunnelling_spectroscopy}a. The dependence on wave vector $k$ and frequency $\omega$ of the spectral function $A(k,\omega)$ could then be determined by analysing the differential tunnelling conductance $\textrm{d}I/\textrm{d}V$. In particular, the authors argued that presence of spin-charge separation should manifest itself as emerging singularities in the $I-V$ characteristics, in a markedly different fashion from what would be expected in a non-interacting system. This idea was later further refined by Grigera \textit{et al.} \cite{grigera_momentum-resolved_2004}, who extended it to arbitrary values of the interaction parameter $K_{\rho}$ while accounting for the effects of Zeeman splitting in both the TLL and the 2DES.

When an electron tunnels quantum-mechanically through a barrier between two regions, it must conserve energy and momentum transverse to the barrier (modulo a reciprocal-lattice vector, in a crystal). It must also obey the Pauli exclusion principle, so it cannot tunnel into an occupied state. In a solid at low temperature, nearly all the states below the Fermi energy \EF are occupied, so this provides a reference energy below which tunnelling is forbidden. We will focus on semiconductor heterostructures where a planar barrier separates two parallel quantum wells containing electrons, which are free to move in two (or fewer) dimensions, so that in those directions their wave functions are described by plane-wave Bloch states with wave vector $\bk$. 

The tunnelling probability then depends on the overlap of the wave functions on either side of the barrier, integrated perpendicular to the barrier, at a rate given by Fermi's golden rule \cite{bardeen_tunnelling_1961,schrieffer_effective_1963,smoliner_tunnelling_1996}. This picks out pairs of initial and final states with matching $\bk$, embodying the conservation of momentum, with resonances in $I$ being observed each time the 2D band and the \textit{n}th subband dispersions are aligned. This, however, is unlikely to happen under equilibrium conditions.

By applying a magnetic field $B$ parallel to the layers, a Lorentz force is produced as the electrons tunnel, boosting the momentum by an amount $\Delta\bk=eBd/\hbar$, where $d$ is the tunnelling distance (the distance between the centres of the wave functions). This allows one to map out the electron dispersion relations on either side of the barrier using the magnetic field  $B$ and the voltage $V_\textrm{dc}$ applied across the barrier, which provides extra energy $eV_\textrm{dc}$. Putting it all together, we obtain
\begin{equation}
\begin{split}
    I\propto & \int \textrm{d}\textbf{k}\textrm{d}E[f_{T}(E-E_{\textrm{F1D}}-eV_{\textrm{dc}})-f_{T}(E-E_{\textrm{F2D}})] \\ & \times A_1(\textbf{k},E)A_2(\textbf{k}+ed(\textbf{n}\times\textbf{B})/\hbar,E-eV_{\textrm{dc}}).
\end{split}
\label{eq:tunnelling_current}
\end{equation}
Here, $e$ is the electronic charge, $f_T(E)$ is the Fermi-Dirac distribution, $d$ is the separation between the wells, $\textbf{n}$ is a unit normal to the plane, $\textbf{B}=-B\hat{\textbf{y}}$ is the (in-plane) magnetic-field vector, and $A_1\left(\textbf{k},E\right)$ and $A_2\left(\textbf{k},E\right)$ are the corresponding spectral functions of the 1D and 2D systems, together with their Fermi energies, $E_{\textrm{F1D}}$ and $E_{\textrm{F2D}}$, respectively (see Fig.\ \ref{fig:tunnelling_spectroscopy}b).

Both the 1999 proposal of Barnes (Altland \textit{et al.} \cite{altland_magnetotunneling_1999}) and the subsequent work of Grigera \textit{et al.} \cite{grigera_momentum-resolved_2004} in 2004 concerned the idea of detecting spin-charge separation in a quantum wire. 
Neither, however, made any attempt at modelling the anomalous suppression of tunnelling current at zero bias, as predicted by the TLL model. This feature, commonly referred to as the zero-bias anomaly (ZBA), is also expected to depend strongly on the strength of the interactions and to follow a power-law behaviour in both bias $V$ and electron temperature $T$.

\begin{figure}[ht!]
  \includegraphics[width=0.7\textwidth]{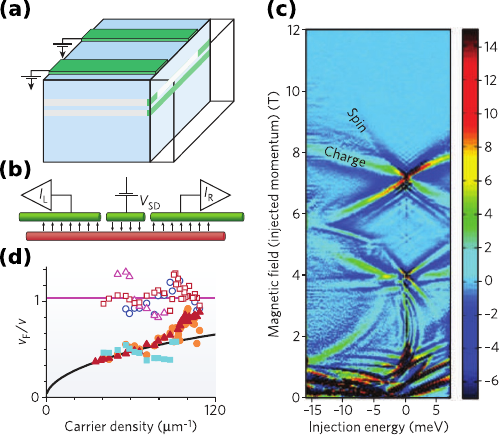}
  \centering
  \caption{Probing spin-charge separation via momentum-resolved magnetotunnelling spectroscopy between two interacting 1D wires. (a) Schematic representation of double-wire structure made via cleaved-edge overgrowth (CEO). The system resides at the edge of a GaAs/AlGaAs double-well heterostructure. (b) Schematic of the circuit used in the 1D--1D tunnelling process. Here, $I_{\textrm{L}}$ is the left-moving current, $I_{\textrm{R}}$ the right-moving current and $V_{\textrm{SD}}$ the source-drain bias voltage. (c) Differential conductance $G=\textrm{d}I/\textrm{d}V$ plotted as a function of both $V_{\textrm{SD}}$ ($\propto \textrm{energy}$) and $B$ ($\propto \textrm{momentum}$). Two features, labelled charge and spin, can be seen branching off at around zero energy and a magnetic field of $7$\,T. (d) Measured spin velocities (open symbols) and charge velocities (filled symbols), plotted as a function of carrier density and normalised with respect to the Fermi velocity $v_{\textrm{F}}$. At lower densities repulsion between the electrons is stronger and hence the charge excitation velocity larger, unlike the spin mode which remains largely unaffected. Solid curves show the theoretical fits. (From \cite{deshpande_electron_2010}, \textit{reprinted with permission}.)}
  \label{fig:1D_1D}
\end{figure}

The first experimental realisation of tunnelling from a 1D wire that was in the single-subband regime was achieved by Auslaender \textit{et al.} \cite{auslaender_tunneling_2002,auslaender_spin-charge_2005}, with later theoretical analysis by Carpentier \textit{et al.} \cite{carpentier_momentum-resolved_2002}. Here, the authors used a pair of coupled quantum wires grown via cleaved-edge overgrowth (CEO), meaning that the tunnelling process took place between two 1D systems instead of between a wire and a 2DES, as discussed before, see Fig.\ \ref{fig:1D_1D}a and b. The CEO technique allowed very strong confinement to be obtained, equal to roughly the width of the well itself, and as such in principle better suited for observing interaction effects. Unfortunately, the probe layer was now also of the same nature as the system being probed, meaning that interpretation and disentanglement of each wire's response became significantly more complicated. These experiments, nevertheless, showed early evidence in support of the TLL picture by observing an enhancement of the excitation velocity relative to the bare electron velocity $v_{\textrm{F}}$, see Fig.\ \ref{fig:1D_1D}c and d. Previous work by the same authors had also observed power-law behaviour of the tunnelling density of states around the zero-bias suppression region \cite{tserkovnyak_finite-size_2002,tserkovnyak_interference_2003}.

In 2009, Jompol \textit{et al.}\ reported the first clear observation of spin-charge separation together with power-law suppression of the tunnelling current at zero bias. In contrast to the CEO samples, the authors fabricated an electrostatically gated 1D system and measured tunnelling between it and a nearby 2DES. Thus the probe layer used was now a well-understood system, making interpretation much more straightforward.

\begin{figure}[ht!]
  \includegraphics[width=0.6\textwidth]{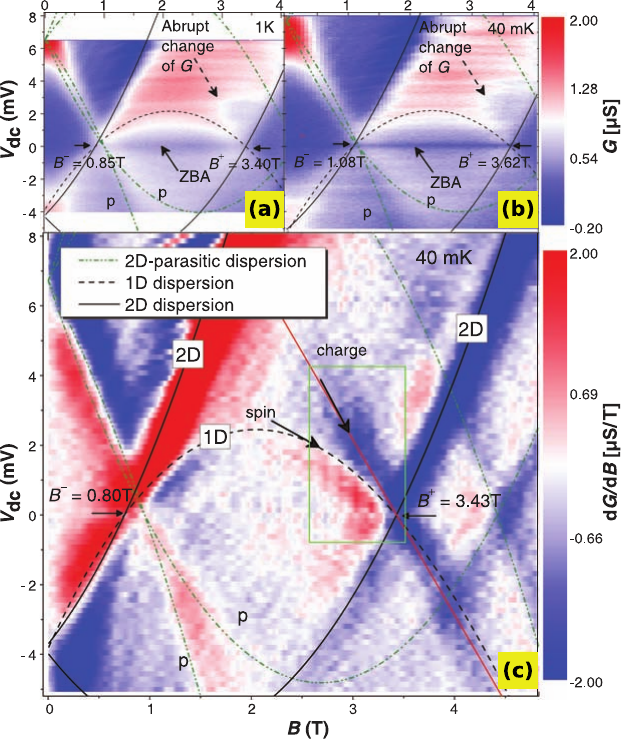}
  \centering
  \caption{Tunnelling conductance $G$ versus dc-bias $V_{\textrm{dc}}$ and magnetic field $B$, at lattice temperatures $T$ of (a) $1$\,K and (b) $40$\,mK. Tunnelling resonances between a system of 1D wires and a 2DES as predicted by the non-interacting model are marked by solid and dashed black lines while green dash-dotted curves indicate the location of unavoidable parasitic 2D--2D tunnelling `p'. In addition, signs of interactions can be seen in the suppression of $G$ around zero bias (ZBA) and its abrupt drop at positive biases around $B^+$. (c) $\textrm{d}G/\textrm{d}B$ differential of the data shown in (b). The non-interacting parabolae shown in (a) and (b) are labelled 1D, 2D, or p, depending on which system's dispersion is being probed. The straight red line, marking the position of the abrupt drop in $G$, is $\sim1.4$ times steeper on this plot than the 1D parabola at $V_{\textrm{dc}}=0$. According to the TLL model, we identify the first as a charge-type mode (holon) while the second as a spin-type excitation (spinon). (From \cite{jompol_probing_2009}, \textit{reprinted with permission}.)}
  \label{fig:conductance_maps}
\end{figure}

Figure\ \ref{fig:conductance_maps} shows the tunnelling conductance $G$ as a function of both inter-layer bias $V_{\textrm{dc}}$ and in-plane magnetic field $B$ at lattice temperatures of 1\,K and $\sim 40\,$mK, when only one 1D subband was occupied. The dashed and solid black curves mark the expected 1D and 2D parabolic dispersion, respectively, arising from single-particle tunnelling in the absence of interactions. A parasitic 2D--2D tunnelling signal originating from the ungated injection region is also shown, as green dash-dotted lines. From the crossing points at $V_{\textrm{dc}}=0$, $B^-$ and $B^+$, the 1D Fermi wavevector $k_{\textrm{F1D}}=ed(B^+-B^-)/2\hbar$ together with the 1D electron density $n_{\textrm{1D}}=2k_{\textrm{F1D}}/\pi$ can be extracted. For the data shown, the approximate electron density in the wires was $n_{\textrm{1D}}\cong 40\,\mu m^{-1}$.

\begin{figure}[ht!]
  \includegraphics[width=\textwidth]{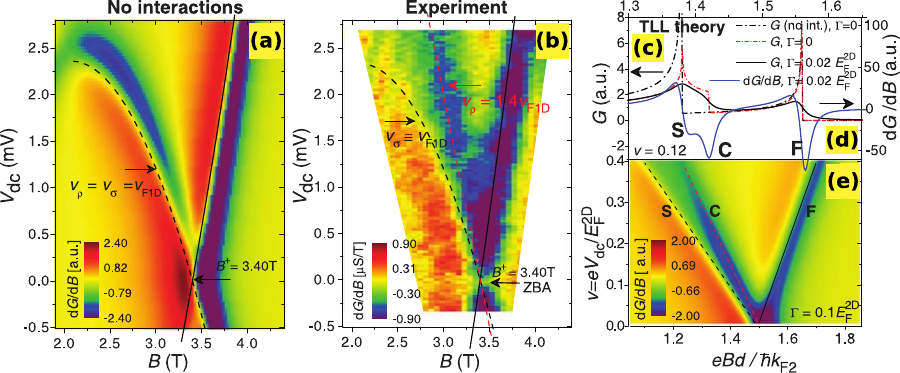}
  \centering
  \caption{Spin-charge separation around the $+k_{\textrm{F}}$ point: comparison between experiment and theory. (a) $\textrm{d}G/\textrm{d}B$ for a system of non-interacting electrons with disorder broadening $\Gamma=0.6$\,meV. (b) High-resolution mapping of $\textrm{d}G/\textrm{d}B$. The red dashed line marks a feature that does not follow the non-interacting parabola (black dashed) and is absent in (a). The spinon velocity is given by $v_\sigma\equiv v_{\textrm{F1D}}$. The extracted holon velocity was $v_\rho=1.4 v_{\textrm{F1D}}$. (c) and (d) show, respectively, the calculation of $G$ and $\textrm{d}G/\textrm{d}I$ for both a non-interacting and a TLL system of electrons. Spin and charge excitations are labeled by S and C, respectively, while F labels the non-interacting 2D dispersion curve. The cuts were taken at a bias of $\nu=eV_{\textrm{dc}}/E_{\textrm{F}}^{\textrm{2D}}=0.12$ (given here in dimensionless units). (e) $\textrm{d}G/\textrm{d}B$ as a function of $B$ and $\nu$ according to the TLL model. Note how, unlike (a), this model predicts a charge feature C as seen in (b). (From \cite{jompol_probing_2009}, \textit{reprinted with permission}.)}
  \label{fig:spin_charge}
\end{figure}

In both Fig.\ \ref{fig:conductance_maps}a and b, an additional region of high conductance at positive biases near $B^+$ can be observed. This can be more clearly seen in Fig.\ \ref{fig:conductance_maps}c by looking at the differential $\textrm{d}G/\textrm{d}B$. In both cases, a feature moving diagonally up and to the left of $B^+$ is visible. Fig.\ \ref{fig:spin_charge}b shows detailed measurements taken around the same region, for a different device, also in the single-subband regime. For comparison, Fig.\ \ref{fig:spin_charge}a and d show the theoretical prediction for a non-interacting system. Here, a strong dark-blue feature of negative $\textrm{d}G/\textrm{d}B$ can be seen, however it follows the 1D parabola unlike what is observed experimentally which moves away from it (dashed red line). It follows therefore that the 1D parabola and the dashed red line track the dispersion of two independent features which, according to the TLL picture, can be associated with spin- and charge-type excitations respectively. The velocity of the charge mode was estimated to be $v_{\rho}=1.4v_{\textrm{F1D}}$, resulting, under the linear approximation, in an interaction parameter $K_{\rho}\approx 0.71$.  For completeness, Fig.\ \ref{fig:spin_charge}c and e show the theoretical result as expected for 1D interacting electrons tunnelling to and from a 2D system according to the TLL model, where a similar feature to that observed can also be seen.

\begin{figure}[ht!]
  \includegraphics[width=0.7\textwidth]{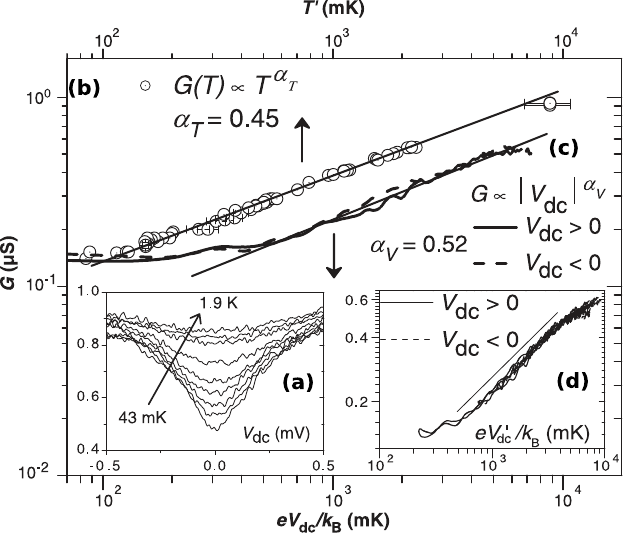}
  \centering
  \caption{The zero-bias anomaly. (a) Tunnelling conductance $G$ as a function of bias $V_{\textrm{dc}}$, taken at a field $B=2.33$\,T, for different temperatures $T$ ranging from 43\,mK to 1.9\,K. (b) Log-log plot of $G(V_{\textrm{dc}}=0)$ versus $T$, showing power-law behaviour $\propto T^\alpha$ with $\alpha_{\textrm{T}}\approx 0.45$. (c) Same as in (b) but showing $G(|V_{\textrm{dc}}|)$ versus $V_{\textrm{dc}}$ with $\alpha_{\textrm{V}}\approx 0.52$. (d) $G(|V_{\textrm{dc}}|)$ versus $V_{\textrm{dc}}^{'}=\sqrt{|V_{\textrm{dc}}|^2+(3k_{\textrm{B}}T/e)^2+V_{\textrm{ac}}^2}$, obtained by adding different sources of energy smearing as noise in quadrature. The value of $3k_{\textrm{B}}T$ was chosen so that the curves in (b) and (c) and all the curves in (d) would superpose. For comparison, using the scaling relation in Eq. (\ref{eq:scaling_conductance}) results in an universal curve with $\alpha\approx0.51$. (From \cite{jompol_probing_2009}, \textit{reprinted with permission}.)}
  \label{fig:ZBA}
\end{figure}

Another feature that cannot be explained by the non-interacting model is the zero-bias anomaly (ZBA), visible as a dark blue horizontal line in Fig.\ \ref{fig:conductance_maps}a and b. As discussed before, this anomalous suppression in the tunnelling current results from interaction effects and is likely to be associated with the extra energy cost for an electron to tunnel in or out of the wire, at zero bias, as it inevitably disturbs the remaining electrons already present. The tunnelling conductance $G$ as a function of $V_{\textrm{dc}}$ for different temperatures $T$ can be seen in Fig.\ \ref{fig:ZBA}a, at a field $B$ approximately midway between $B^-$ and $B^+$. Similarly, Fig.\ \ref{fig:ZBA}b and c show $G(V_{\textrm{dc}}=0,T)$ and $G(|V_{\textrm{dc}}|,T<70$\,mK), respectively, on a log-log plot. Here, clear variation as a power law in both $V_{\textrm{dc}}$ and $T$ over a range of orders of magnitude is observed. Both $V_\textrm{dc}$ and $T$ play a similar role in smearing the energy, as illustrated in Fig.\ \ref{fig:ZBA}d.

The interaction parameter $K_{\rho}$ was obtained from the extracted power-law exponents, $\alpha_{\textrm{T}}\approx 0.45\pm0.04$ and $\alpha_{\textrm{V}}\approx 0.52\pm0.04$, depending on whether excitations from electrons tunnelling into or out of the wire generally propagated as far as the ends (`end tunnelling') or not (bulk tunnelling'), see section \ref{sec.transport}. For the latter, the measured exponents gave $K_{\rho}\approx0.28$ and 0.26 respectively, significantly smaller than the value extracted from the holon branch in Fig.\ \ref{fig:spin_charge}b. By considering `end' tunnelling instead, the values obtained were of $K_{\rho}=0.53$ for $\alpha_{\textrm{T}}$ and $K_{\rho}=0.49$ for $\alpha_V$, considerably closer to that obtained in Fig.\ \ref{fig:spin_charge}. However, this implies that the wire length is shorter than the thermal length $L_T= 2\hbar v_\textrm{F1D}/(K_{\rho}k_\textrm{B}T)$ at all temperatures measured (100\,mK to 10\,K), or else the power-law exponent would change at some temperature, as observed in 1D--1D tunnelling by Tserkovnyak \textit{et al.} \cite{tserkovnyak_interference_2003}.

Overall, the work carried out by Jompol \textit{et al.}\ showed clear evidence in support of the TLL model. It was obvious, though, that the `charge' line observed persisted far beyond the energy range over which the dispersions could be approximated to be linear, making the spin-charge separation a phenomenon that persisted beyond the range of validity of the TLL model. This raised the question, therefore, of what happens to both spin and charge modes in the nonlinear region, that is, far from the Fermi points, where the linear approximation breaks down. The role of interactions in 1D wires at arbitrary energy and momentum has only recently started to be understood. Such `nonlinear Tomonaga-Luttinger liquids' are the topic of the next section.

\section{Beyond the linear Tomonaga-Luttinger-liquid approximation}

TLL theory works remarkably well in describing the low-energy behaviour of one-dimensional systems. As  we have shown, these exhibit a number of unusual properties, including long-lived plasmonic excitations, the separation of the spin and charge degrees of freedom, and interaction-driven power-law behaviour \cite{haldane_textquotesingleluttinger_1981}. In general, however, the 1D spectrum is curved, \emph{i.e.} nonlinear, meaning that deviations from linearity are expected to occur away from the Fermi points. We will now focus our attention on two particular nonlinear theories. 

\subsection{Mobile-impurity model of nonlinear Tomonaga-Luttinger liquids}

\begin{figure}[h]
  \includegraphics[width=0.8\textwidth]{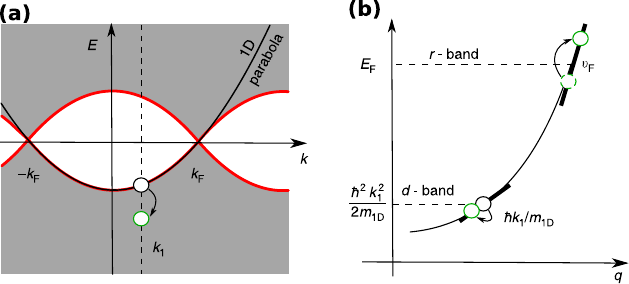}
  \centering
  \caption{(a) Dispersion of an interacting 1D system. Kinematically forbidden regions are shown in white while the grey shaded area represents the many-body continuum of excitations. The spectral edge marks the border between these two regions and is shown in red. A higher-energy excitation (green circle) is  obtained by changing the state of a heavy hole (black circle) deep below the Fermi energy and simultaneously creating some linear TLL excitations around the Fermi energy. This moves the total energy of the many-body state away from the spectral edge as described by the mobile-impurity model---see text. (b) Splitting of the fermionic dispersion into two subbands, one for a heavy hole deep in the Fermi sea, with characteristic velocity given by $\hbar k_1/m_{\textrm{1D}}$, and one for excitations close to the Fermi point $k_{\textrm{F}}$, with velocity $v_{\textrm{F}}$. (From \cite{jin_momentum-dependent_2019}, \textit{reprinted with permission}.)}
  \label{fig:Jin_fig2_adapted}
\end{figure}

Excitations away from the Fermi energy are described in a one-dimensional system by nonlinear hydrodynamics, in what came to be known as the mobile-impurity model, see Fig.\ \ref{fig:Jin_fig2_adapted} \cite{imambekov_universal_2009,imambekov_one-dimensional_2012}. The name itself is reminiscent of another problem, X-ray scattering in metals, where a hole is created
far below the Fermi level upon absorption of a high-energy X-ray photon. As in the TLL model, this system is also known to have power-law singularities around $E_{\textrm{F}}$. However, in contrast to that model, perturbation analysis on the resulting interaction between the deep hole and the low-energy quasiparticles is divergent. 

Nozi\`{e}res and De Dominicis\cite{nozieres_singularities_1969} solved this problem by introducing the so-called heavy-impurity model. This construction, resulting from Fermi-liquid quasiparticles interacting with a hole state deep below the Fermi level, can be diagonalised exactly, in a framework analogous to that of a TLL system. It also predicts power-law behaviour away from the Fermi energy, in the form of a threshold exponent. Most interestingly though, the 1D spectral function, as derived from the mobile-impurity model, is
\begin{equation}
A_1(k_x,E)\propto\frac{1}{|E+\mu-\xi_1|^{\alpha(k_x)}},
\label{eq:mobile_impurity}
\end{equation}
implying that there is now a momentum dependence in the exponent $\alpha(k_x)$ of the interacting 1D spectral function\cite{ImambekovRMP12}. The explicit expression for spinful fermions that is applicable for electronic systems is found to be \cite{TsyplyatyevPRB14} 
\begin{equation}
    \alpha(k_x)=1-\frac{\left(1-C(k_x)\right)^2}{4K_{\rho}}-\frac{K_{\rho}\left(1-D(k_x)\right)^2}{4},
\end{equation}
where $C(k)=\left(k^2-k_{\textrm{F}}^2\right)/\left(k^2/K_{\sigma}-k_{\textrm{F}}^2K_{\sigma}/K_{\rho}^2\right)$ and $D(k)=\left(k-k_{\textrm{F}}\right)\left(k_{\textrm{F}}/K_{\rho}^2+k/K_{\sigma}^2\right)/$\hfill$(k^2/K_{\sigma}^2-k_{\textrm{F}}^2/K_{\rho}^2)$.

Having now extended the universality of a TLL from the low- to the finite-energy regime, one finds that a momentum-dependent power-law emerges due to the finite curvature of the spectral edge in energy and momentum space. This is a unique feature of nonlinear hydrodynamics in 1D and a hallmark of nonlinear Tomonaga-Luttinger-liquid (NLL) behaviour. Another implication of this theory is the emergence of the `replicas' of the principal dispersion, \emph{i.e.}  all red lines in Fig.\ \ref{fig:Jin_fig2_adapted}a are predicted to be pronounced in the spectral function instead of just the one coinciding with the principal 1D parabola.

Yet another prediction from the NLL model is the drastic reduction of the plasmon's lifetime. This can be understood by analogy with the linear case. If the spectrum is linear, each electron-hole pair can be identifiable by an energy linear to its momentum. If, instead, the spectrum is curved, the possibility exists of having different electron-hole pairs with different energies yet the same momentum. These can morph into a plasmon with wave number $q$ once the interactions are turned on, with a finite spectral width, approximately equal to the spread in energies, $\delta\omega(q)=\omega_+(q)-\omega_-(q)$, and hence a reduced lifetime $\tau\propto 1/\delta\omega$.

\subsection{A Mode-hierarchy away from the Fermi points}

The mobile-impurity model, described in the previous section, has released the TLL theory from the restriction of the linear approximation. Nevertheless, it still does not provide a description for arbitrary nonlinear excitations since it is applicable only in the proximity of the spectral threshold. This is because, being constructed by analogy with the edge singularity in the X-Ray scattering problem, the nonlinear hydrodynamics become less well-defined away from the spectral edge separating the forbidden zone from the continuum of many-body excitations in 1D, where such a phenomenological approach becomes inconsistent. 

\begin{figure}[h]
  \includegraphics[width=0.9\textwidth]{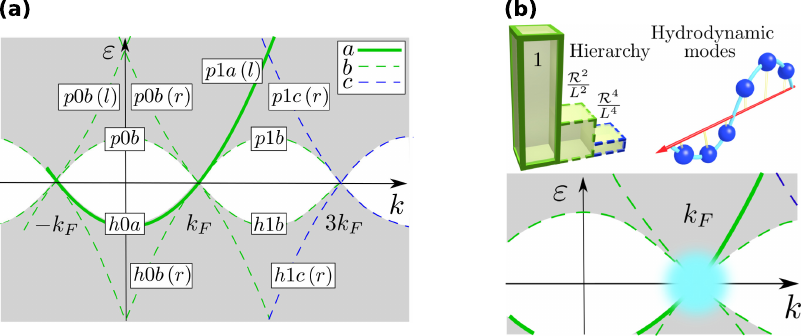}
  \centering
  \caption{(a) Spectral function of spinless fermions according to the mode-hierarchy picture. First (second)-level modes are shown in the region $-k_{\textrm{F}}<k<k_{\textrm{F}}$ ($k_{\textrm{F}}<k<3k_{\textrm{F}}$) and are labelled by 0 (1), where $k_{\textrm{F}}$ is the Fermi wave vector. Accessible and forbidden regions are marked by grey and white, respectively. Particle (hole) sectors are marked by $p$ ($h$) for positive (negative) energies, $a$, $b$ and $c$ correspond respectively to the level in the mode-hierarchy in powers of 0, 1 and 2 of $R^2/L^2$ and $(r,l)$ correspond to the origin in the range, right or left. (From \cite{tsyplyatyev_hierarchy_2015}, \textit{reprinted with permission}.) (b) Bottom, regions of validity for different theories on the energy-momentum plane. Top right, TLL hydrodynamic modes, where the deviations from the red line mark a variation in density and which dominate at low energies (see cyan region in the bottom panel). Top left, mode-hierarchy up to third order, where the bars denote qualitatively the amplitude of different many-body modes and which is relevant for the rest of the plane. (From \cite{tsyplyatyev_nature_2016}, \textit{reprinted with permission}.)}
  \label{fig:tsyplyatyev_hierarchy}
\end{figure}

A more systematic understanding of the general picture for nonlinear excitations came from a microscopic analysis via the Bethe ansatz. It was found \cite{tsyplyatyev_hierarchy_2015,tsyplyatyev_nature_2016} that an exponentially large number of excitations are separated into levels of a mode-hierarchy according to their spectral strength, which is proportional to integer powers of a small parameter $R^{2}/L^{2}$, where $R$ is the radius of the two-body interaction potential and $L$ is the length of the system. A detailed analysis of these amplitudes revealed formation of a parabola-like dispersion, see Fig.\ \ref{fig:tsyplyatyev_hierarchy}a, similar to that of original non-interacting fermions but with some renormalisation due to interactions, by the strongest excitations with the zeroth power of $R^{2}/L^{2}$. All other many-body modes have powers of $R^{2}/L^{2}$ greater than zero with the general trend of increasing the integer by one with each discrete step of $2k_{\textrm{F}}$ away from the principal parabola.

As an example, the hole part of the principal parabola, between the $\pm k_{\textrm{F}}$ points, has the largest amplitude but its mirror in the particle sector---a `replica' in the shape of a dome marked as $p0b$ in Fig.\ \ref{fig:tsyplyatyev_hierarchy}a---has a parametrically smaller amplitude proportional to the first power of $R^{2}/L^{2}$. In the spectral function the strength of this replica is predicted to be 
\begin{equation}
A_{1}\left(k_{x},E\right)\propto\frac{R^{2}}{L^{2}}\frac{k_{\textrm{F}}^{2}k_{x}^{2}}{\left(k^{2}-k_{\textrm{F}}^{2}\right)^{2}}\delta\left(E-\mu+\xi_{1}\right).
\end{equation}
Therefore, for almost all momenta this mode will be unobservable in the thermodynamic limit. The only exceptions are the regions around the $\pm k_{\textrm{F}}$ points, where the singularity in the denominator starts to compete with the parametric smallness, resulting in a large amplitude overall. On the other hand, the measurement of the whole mode requires using smaller systems, in which the $R^{2}/L^{2}$ parameter still leaves the amplitude of the whole mode above the background from the other processes.

Close to the spectral threshold the mode-hierarchy reproduces the predictions of the mobile-impurity model. The small parameter $R^{2}/L^{2}$ provides a path for accounting for the principal amplitudes of the exponential continuum of many-body excitations, which leads to the microscopic calculation of the spectral function. The resulting threshold exponents match those predicted by the phenomenologically introduced nonlinear hydrodynamic model.

While the mode-hierarchy emerges away from the Fermi points, close to them (where the spectrum is almost linear) it transitions into the usual linear TLL, see Fig.\ \ref{fig:tsyplyatyev_hierarchy}b. The hydrodynamic modes of the latter consist of a huge number of many-body modes, all of which are of similar spectral strengths, making the two regimes distinct already on the microscopic level. The change from one into the other can be traced quite easily using a macroscopic quantity, the density of states. It can be calculated exactly by numerical means using Bethe ansatz, exhibiting the power-law suppression around the Fermi energy $E_{\textrm{F}}$ predicted by the TLL model and a crossover into a finite density $\propto1/\sqrt{E}$ predicted by the mode-hierarchy away from the linear region, where the nonlinearity of the single-particle dispersion already destroys the hydrodynamic modes of the TLL.

\newpage

\section{Recent work on nonlinear effects}

\subsection{1D `replica' modes}

\begin{figure}[ht]
  \includegraphics[width=0.95\textwidth]{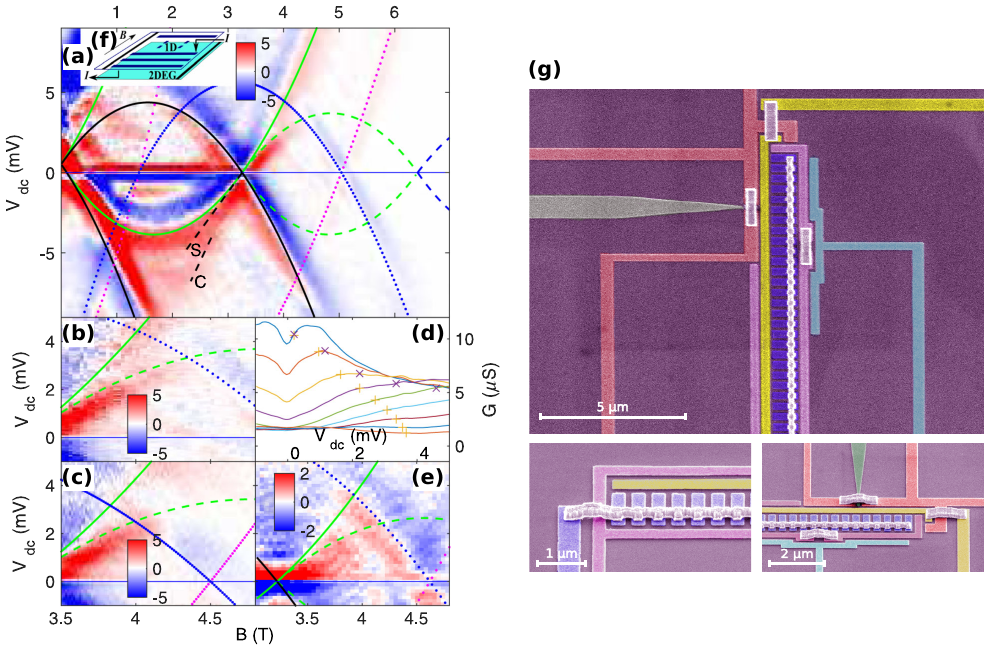}
  \centering
  \caption{Tunnelling differential conductance $G=\textrm{d}I/\textrm{d}V$ for two samples, consisting of a set of identical wires of length $L=10\,\mu$m (a)--(d) and $L=18\,\mu$m (e). A schematic representation of the device measured can be seen in inset (f). (a) $\textrm{d}G/\textrm{d}V_{\textrm{dc}}$ map obtained at a lattice temperature of $T=300$\,mK. The 1D signal is marked by solid green curves for the $a$ modes, dashed green for the $b$ modes and dashed blue for the $c$ modes, see Fig.\ \ref{fig:tsyplyatyev_hierarchy} for details. Magenta and blue curves indicate the parasitic 2D--2D signal while the black curve follows the dispersion of the 2D system as probed by the 1D wires. Spin and charge modes are labelled by the S and C lines, respectively. (b) and (c) show an enlargement of the replica region to the right of $+k_{\textrm{F}}$, as seen in (a), for $v_{\textrm{PG}}>0$ and $v_{\textrm{PG}}=0$, respectively, where PG is a gate running over most of the parasitic `p' region, see blue curves. (d) $G$ vs $V_{\textrm{dc}}$ at various fields $B$, around the replica region. The symbols $+$ and $\times$ on each curve mark, respectively, the dashed and solid green curves in (a) and (b). (e) $\textrm{d}G/\textrm{d}V_{\textrm{dc}}$ map for a second device obtained at $T<100$\,mK, showing a similar feature to that observed in (a)--(c). (From \cite{tsyplyatyev_hierarchy_2015} \textit{reprinted with permission}.) (g) Scanning electron micrographs of a tunnelling device with air-bridges, showing how air-bridges are used not only to connect gates together across other gates, but also to link together all the finger gates defining the 1D wires, which avoids variation of the gate potential along each of these very short wires, as happens when the gates are joined together at one end \cite{jin_microscopic_2021,Vianez_2021}}
  \label{figc:tsyplyatyev_fig3}
\end{figure}

In 2015, the first observation of structure resembling higher-order excitations, as predicted by the mode-hierarchy model, was reported in Tsyplyatyev \textit{et al.} \cite{tsyplyatyev_hierarchy_2015} in devices similar to that used by Jompol \textit{et al.} \cite{jompol_probing_2009} but with air-bridges \cite{jin_microscopic_2021}, see Fig.\ \ref{figc:tsyplyatyev_fig3}. The experiment consisted of measuring momentum-resolved tunnelling to and from an array of one-dimensional wires to a 2DES in a GaAs/AlGaAs heterostructure, for a variety of different wire lengths. The authors observed structure resembling second-level spinon excitations, near the $+k_{\textrm{F}}$ point, branching away from the 1D mode and dying away rapidly at high momentum, in line with the theoretical predictions. Most interestingly, however, this feature was observed to emerge as a function of system length, its spectral weight increasing for shorter channels. It was also noted to be visible even when more subbands started being occupied. A comprehensive review on the mode-hierarchy picture, highlighting both theoretical and experimental progress, can be found in Tsyplyatyev \textit{et al.} \cite{tsyplyatyev_nature_2016}.

\begin{figure}[ht]
  \includegraphics[width=0.9\textwidth]{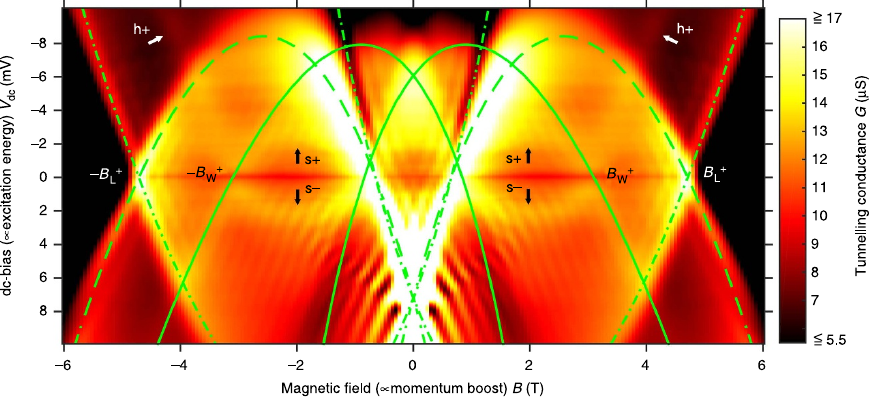}
  \centering
  \caption{Tunnelling conductance $G$ plotted as a function of dc-bias $V_{\textrm{dc}}$ and in-plane magnetic field $B$, for the regime of a single 1D subband occupied. The electron density in the wires is $n_{\textrm{1D}}\approx35\,\mu m^{-1}$. The solid green curves mark the dispersion of the 2D system as mapped by the 1D wires, while the green dashed and dash-dotted lines indicate the resonances arising from 2D--2D parasitic tunnelling. Spinon bands can be seen in the hole (s--) and particle (s+) sectors, while a holon band is present in the particle sector (h+). The labels $\pm B_{\textrm{W,\textrm{L}}}^+$ indicate specific magnetic fields at which each resonances crosses the $V_{\textrm{dc}}=0$ axis. (From \cite{moreno_nonlinear_2016}, \textit{reprinted with permission}.)}
  \label{fig:moreno_fig5}
\end{figure}

In 2016, Moreno \textit{et al.} \cite{moreno_nonlinear_2016}, also using tunnelling spectroscopy, found evidence for the existence of another higher-order mode, this time a first-level inverted spinon band (see Fig.\ \ref{fig:moreno_fig5}). The tunnelling device used consisted of an array of 6000 GaAs wires, $1\,\mu$m in length, considerably shorter than previously used in Tsyplyatyev \textit{et al.} \cite{tsyplyatyev_hierarchy_2015,tsyplyatyev_nature_2016}, which were $10\,\mu$m and $18\,\mu$m in length. The mode reported by Moreno \textit{et al.} could not be observed in the longer-length wires, further establishing the mode-hierarchy picture as one of the main nonlinear TLL theories. More recently, a full experimental study of the length-dependence of replicas has been carried out by Vianez \textit{et al.} \cite{Vianez_2021}, confirming the prediction in the mode-hierarchy picture that the replicas get stronger as the length decreases \cite{tsyplyatyev_hierarchy_2015,tsyplyatyev_nature_2016}.

\subsection{Momentum-dependent power law}

As was discussed in the previous section, the Mobile-Impurity model, developed in the context of 1D quantum fluids at arbitrary energy and momentum, makes a remarkable prediction regarding the power-law behaviour of the observed density of states near the spectral edge. Here, due to the finite curvature of the 1D dispersion, the exponent becomes momentum dependent, in clear contrast with the linear model, where no such dependence is expected. For electrons, \emph{i.e.} spin-1/2 fermions, the observation of a momentum dependence of the threshold exponent away from the Fermi points would constitute a hallmark of 1D nonlinear hydrodynamics.

In 2019 Jin \textit{et al.}  \cite{jin_momentum-dependent_2019} reported the first observation of this new type of power-law using a system of interacting electrons in a quantum wire, probed via tunnelling spectroscopy. Here, a finite length-scale was present---given by the particle's wavelength---unlike in the usual physical picture of phase transitions where length-scales become infinite, making it the first power-law observed in the absence of a scale invariance. 

\begin{figure}[ht]
  \includegraphics[width=0.9\textwidth]{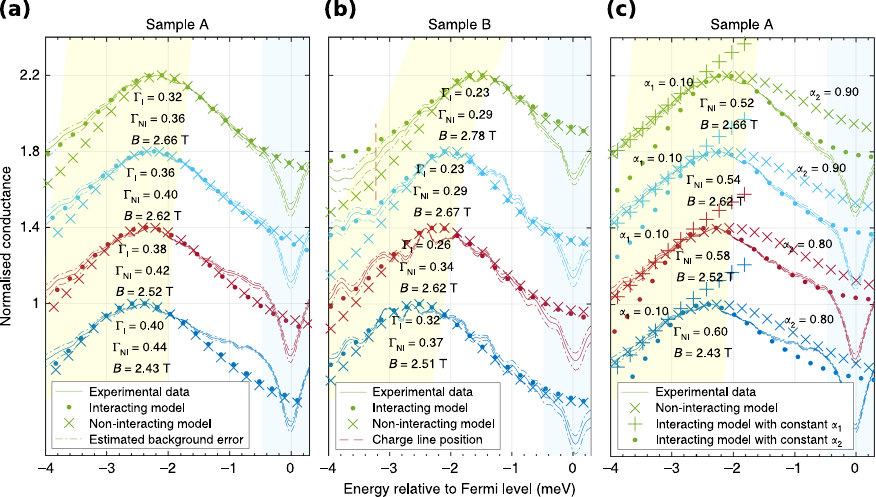}
  \centering
  \caption{Momentum-dependent power law. (a) and (b) Fits to the tunnelling conductance of two different samples, around the bottom of the 1D subband, for a variety of different fields, normalised to their peak value and shifted vertically for clarity. Two different models, non-interacting ($\times$) and interacting with a momentum-dependent exponent ($\cdot$), were used when fitting the data. The effects of disorder-induced broadening were also considered. Sample A, consisting of $18\,\mu m$ long wires, was measured at $50$\,mK in a $\textrm{He}^3/\textrm{He}^4$ dilution refrigerator, while sample B, with $10\,\mu m$, was measured at $330$\,mK in a $\textrm{He}^3$ cryostat. (c) Fittings obtained using a momentum independent power law ($\cdot$ and $+$) and a non-interacting model ($\times$), matched to specific parts of the data. (From \cite{jin_momentum-dependent_2019}, \textit{reprinted with permission}.)}
  \label{fig:Jin_momentum}
\end{figure}

The authors observed that an enhancement of the tunnelling conductance below the bottom of the 1D dispersion (see Fig.\ \ref{fig:Jin_momentum}) could not be explained solely by considering either a non-interacting or linear Luttinger framework (\emph{i.e.} a momentum-independent power law). Instead, the best fits to the data were obtained by considering a momentum-dependent power law as predicted by the mobile-impurity model. In addition to the interaction effects, the effects of disorder-induced broadening were also considered. This result was observed in multiple devices, with different wire lengths and measured at different temperatures.

The extracted Luttinger parameter from the nonlinear regime was $K_{\rho}=0.70\pm0.03$. This is in good agreement with the values obtained from other 1D interaction effects, most notably the ZBA ($K_{\rho}=0.59\pm0.13$ assuming end-tunnelling) and the spin-charge separation ($K_{\rho}=0.76\pm0.07$), both also present in this experiment. Note that throughout this analysis the authors considered $K_{\sigma}=1$, in accordance to renormalisation-group arguments \cite{giamarchi_quantum_2003}.

\subsection{Lifetime of spinon and holons at high energies}

One of the early studies of nonlinear TLL dynamics was done by Barack \textit{et al.} \cite{barak_interacting_2010}. Here, the authors selectively injected holes and particles into a quantum wire in order to study its relaxation properties via momentum-resolved tunnelling spectroscopy. Their measurements concluded that while energetic particles underwent a rapid thermal relaxation, holes remained largely unchanged, in clear contrast with the linear TLL theory, in which energy relaxation is largely forbidden. This set the limits on the relaxation times to $\tau<10^{-11}$\,s for particles and $\tau\gg10^{-11}$\,s for holes. 

It is important to highlight that these results rest on the assumption that the interactions are weak. Previous work done by the same authors on similar samples obtained a Luttinger parameter for the electrons of $K_\rho\approx0.55$, comfortably sitting in the strongly interacting regime. This apparent contradiction however was explained by suggesting that relaxation dynamics are largely independent of the strength of the interactions. The model derived showed good agreement with the data. Its phenomenological nature however illustrated the clear need for a microscopic description of kinetics and energy relaxation in the strongly interacting limit.

\subsection{Nonlinear carbon nanotubes}

Recent work by Wang \textit{et al.} \cite{wang_nonlinear_2020} has provided further experimental evidence of nonlinear TLL dynamics beyond the limitations observed in Barak \textit{et al.} By growing long and clean semiconducting carbon nanotubes (CNTs), the authors were able to utilise low-noise near-field scanning optical measurements to map the plasmonic excitations in a nanotube, even in the low-density limit. Semiconducting CNTs have a bandgap and as such the linear approximation is expected to break as one approaches the bottom of the conduction band. By coupling the nanotube to a gate electrode, the authors were able to tune the position of the Fermi level and use it to probe the effect of the nonlinear dispersion on the plasmons. Similar measurements were also carried out on metallic CNTs, where the band structure is gapless and so captured well by the linear model.

\begin{figure}[ht]
  \includegraphics[width=0.9\textwidth]{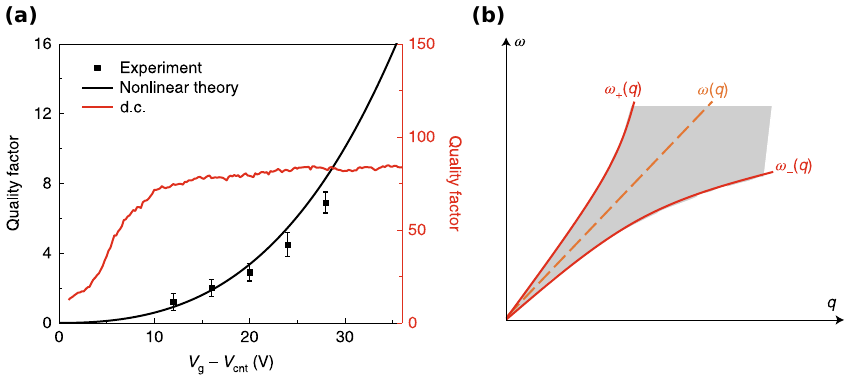}
  \centering
  \caption{Nonlinear Luttinger plasmons in carbon nanotubes. (a) Plasmon quality factor as a function of gate voltage. The experimental results match the NLL well (black) and cannot be captured by alternative mechanisms like impurity scattering (red). (b) DSF diagram characteristic of the NLL model. In a nonlinear TLL, the plasmon mode is not an exact eigenstate of $\omega(q)=v_{\textrm{p}}q$ (dashed) but instead is broadened with upper and lower bounds given by $\omega_\pm(q)$, respectively. The spectral width, $\delta\omega(q)=\omega_+(q)-\omega_-(q)$, therefore increases with $q$, indicating a reduction in the plasmon lifetime, as observed. (From \cite{wang_nonlinear_2020}, \textit{reprinted with permission}.)}
  \label{fig:Wang_nonlinear}
\end{figure}

One of the key predictions of the NLL theory is a drastic reduction of the plasmon lifetime. By injecting a plasmon wave into the nanotube, Wang and collaborators were able to observe the interference between the incident and reflected excitations at one end of the nanotube and from there extracted the quality factor and the wavelength of the mode in the system. This allowed them to measure the increase of the quality factor (and consequently, of the plasmon lifetime) as a function of the electron density, see Fig.\ \ref{fig:Wang_nonlinear}a. By contrast, the linear model predicts infinitely lived excitations, with no intrinsic relaxation mechanism and therefore gate-independent. Optical and transport measurements on similar samples also confirmed that the decay was indeed associated with nonlinear dynamics rather than scattering of impurities.

Their results can be understood from the dynamic structure factor (DSF) of a NLL, see Fig.\ \ref{fig:Wang_nonlinear}b. For a linear TLL, such as in a metallic nanotube, the DSF takes the form $S(q,\omega)=2K|q|\delta(\omega-v_{\textrm{p}}q)$. This results in a linear dispersion for the plasmon mode, given by $\omega(q)=v_{\textrm{p}}q$. Here, $K$ is the Luttinger parameter, $v_{\textrm{p}}$ is the plasmon velocity (charge mode), and $q$ is its wave vector.

In contrast, for a nonlinear TLL, like in a semiconducting CNT, the DSF is given by
\begin{equation}
    S(q,\omega)=2\frac{\tilde mK}{|q|}\theta\left(\frac{q^2}{2\tilde m}-|\omega-v_{\textrm{p}}q|\right),
\end{equation}
where $\tilde{m}$ is an interaction-dependent effective mass and $\theta$ is the Heaviside function.

This means that the plasmon mode is no longer an exact eigenstate of the system but instead is broadened, with upper and lower limits given by $\omega_{\pm}=v_{\textrm{p}}q\pm q^2/\left(2m\right)$. Consequently, for a given frequency, \emph{i.e.}\ energy, there will now be multiple available plasmons with different momenta. A high-energy plasmon can then easily decay into multiple low-energy plasmons. The damping and lifetime associated with this process is determined by the width of the broadening $\delta\omega(q)=q^2/\tilde m$.

\section{Other work on 1D interaction effects}

\subsection{Coulomb drag}

Free carriers in semiconducting devices screen each other weakly and so the Coulomb interaction between charges is screened over a finite distance, which is however often rather long. This means that when two parallel conducting wires come together, separated only by a thin insulating barrier, a current through one of them can induce a net charge displacement in the other. This is called Coulomb drag. The drag resistivity $R_{\textrm{D}}=-V_{\textrm{drag}}/I_{\textrm{drive}}$ is determined by the drive current in the inducing wire, $I_{\textrm{drive}}$, and the measured drag voltage in the induced wire, $V_{\textrm{drag}}$.

Initially, it was thought that this phenomenon was due to electrons in the drive wire transferring their momentum to their counterparts in the other wire, via the Coulomb interaction, keeping the original direction of motion unchanged. However, recent work by Laroche \textit{et al.}  \cite{laroche_towards_2008,laroche_positive_2011} strongly challenged this picture. They found that the drag resistance could be both positive and negative, while the momentum-transfer model can only account for the former. Negative Coulomb drag, however, has been predicted to exist in models where the drag is induced by charge fluctuations, therefore introducing a new transport paradigm in addition to the momentum-transfer picture.

Later work by the same group  \cite{laroche_1d-1d_2014} also observed an upturn in drag resistance at a temperature $T^\star$ when in the single-subband regime. This confirmed a long-held prediction of the TLL model, wherein a crossover between drag dominated by forward (for $T>T^\star$) or backward scattering (for $T<T^\star$) should occur, thus establishing the 1D--1D drag dependence on temperature. For comparison, the conductance along a 1D channel in the ballistic regime has only a very weak dependence on $T$.

\subsection{Helical current}

Conductance along a ballistic, one-dimensional, quantum wire is quantised in units of $2e^2/h$, where the factor of 2 arises from the spin degeneracy of the electrons. This is not affected by interactions in a clean system, as it is solely determined by the contact resistance to the Fermi leads. In the presence of disorder, however, it is reduced following a TLL power law. 

Recent work carried out in the ultra-low-temperature regime ($T\sim10$\,mK) in CEO wires revealed promising new evidence for helical nuclear magnetism in the TLL regime. Tracking the conductance of the first mode as a function of temperature, Scheller \textit{et al.} \cite{scheller_possible_2014} observed the expected value of $2e^2/h$ at $T\gtrsim 10$\,K drop to $e^2/h$ as the temperature was lowered, becoming independent of it at $T\lesssim0.1$\,K. This behaviour was seen to be independent of both density and magnetic field up to at least 3\,T. 

After ruling out several other potential interpretations, the authors concluded that the most likely explanation is the appearance of a nuclear spin helix---an ordered state of nuclear spins hosted by atoms of the same crystal---thus explaining the drop in conductance by a factor of 2 at lower temperatures, since the wire now only transmits spin-down right- and spin-up left-movers. This is consistent with a recent theory proposed by Braunecker \textit{et al.} \cite{braunecker_nuclear_2009-1,braunecker_nuclear_2009,braunecker_spin-selective_2010}.

\subsection{Cold atoms}

For completeness, we conclude by commenting on a different platform, outside solid state, for studying the physics of strongly correlated many-body systems. In recent years, experiments have emerged reporting the formation of clean, disorder-free 1D chains by trapping ultracold atoms in optical lattices \cite{pagano_one-dimensional_2014,boll_spin-_2016,yang_measurement_2018}. Here, the confinement potential is given by the laser field itself, therefore allowing the simulation of a very controlled Fermi-Hubbard model not otherwise accessible to material systems. Recent work using quantum gas microscopy, that is, probing the atomic chain with single-atom/single-lattice-site resolution, has revealed signatures of spin-charge separation \cite{hilker_revealing_2017-1,salomon_direct_2019}, with real-space tracking of the deconfinement and evolution of each separate plasmon having also been reported \cite{vijayan_time-resolved_2020}. These systems are particularly attractive because of the relative ease of varying the strength of the interactions within them, and because they also offer the prospect for tracking crossover from 1D to 2D physics.

\vspace{1in}
\section{Conclusion}
Tunnelling spectroscopy has proven itself over the past years as a powerful technique capable of mapping the dynamics of many-body systems across the whole spectrum of momentum and energy. This has been shown to be extremely important, for not only has it allowed the confirmation of many of the predictions made by the original, linear TLL model, but also, more recently, it has provided much-needed evidence in support of its nonlinear counterparts. Improvements in wafer quality, nanofabrication and tunnelling resolution have also recently allowed for a full mapping to occur at high energies, with early results suggesting signs of physics beyond the Tomonaga-Luttinger paradigm. We anticipate that, provided that two conductors lie close together with a well-controlled tunnel barrier in between, this technique can be applied to study different classes of materials, such as topological insulators, therefore offering itself as a powerful tool for probing exotic single-particle and many-body physics in new systems.

\bibliography{scibib.bib}

\bibliographystyle{Science}

\end{document}